\documentstyle[aps,eqsecnum,preprint]{revtex} \baselineskip 12pt
\begin{document}
\title{A Number of Quasi-Exactly Solvable $N$-body Problems}
\preprint{9803003}

\vspace{.4in}

\author{Avinash Khare }

\address{
Institute of Physics, Sachivalaya Marg,\\ Bhubaneswar-751005, India,\\
Email:  khare@iopb.stpbh.soft.net}
\author{Bhabani Prasad Mandal}
\address{
Theory Group, Saha Institute of Nuclear Physics,\\ 1/AF, Bidhannagar
Calcutta-700064, India,\\
Email: bpm@tnp.saha.ernet.in}
\vspace{.4in}

\maketitle

\vspace{.4in}

\begin{abstract}

We present several examples of quasi-exactly solvable $N$-body problems in one, two and higher 
dimensions. We study various aspects of these problems in some detail. In 
particular, we show that in some of these examples the corresponding polynomials
form an orthogonal set and many of their properties are similar to those of
the Bender-Dunne polynomials. We also discuss QES problems where the
polynomials do not form an orthogonal set. 

\end{abstract}

\newpage
\section{Introduction}
 
In last few years, the quasi-exactly solvable (QES) problems have attracted 
a lot of attention \cite{ush}. In particular, by now a detailed study has 
been made of several QES problems in (one particle) non-relativistic quantum 
mechanics in one dimension and several interesting features have been 
uncovered. However, very few QES problems in multi-dimensions or 
$N$-body problems in one dimension \cite{ush1,mrt} have been discussed so far. 
Further, to
the best of our knowledge, no QES $N$-body problem in two and higher dimensions
has been discussed so far. The purpose of this note is to initiate a 
systematic study of the various aspects of the $N$-body QES problems in
multi-dimensions. In particular, one would first like to discover several
$N$-body QES problems in two and higher dimensions. Further, one 
would like
to know if the corresponding polynomials form an orthogonal set or not. 
Besides, one would like to know if there is an underlying hidden algebra.
In this context, it is worth recalling that in the case of the one body problem 
in one dimension, it is known that in most QES cases (though not all 
\cite{jat}), there is 
an underlying hidden algebra $Sl(2)$. Further,   
the polynomials form an orthogonal set
provided the Hamiltonian can be written in terms of the quadratic generators
of this $Sl(2)$ algebra \cite{bpm} and that in this case one has one 
\cite{bd} or two
sets \cite{bpm1} of Bender-Dunne polynomials.   

In this paper we discuss in some detail three different $N$-body QES problems
namely, Calogero-Marchioro type $N$-body problem in D-dimensions \cite{cm,kr}
in Sec.II ($D \ge 2$),
$N$-body system exhibiting novel correlations in two dimensions \cite{mbs}
in Sec. IIIA, 
and Calogero-Sutherland type $N$-body problem in one dimension \cite{cal} 
in Sec. IIIB. 
In each case
we discuss QES problem in case the potential is either of the sextic-type or
is a sum of oscillator plus Coulomb type-potential. We show that whereas in
the former case, the polynomials form an orthogonal set, in the later case they
do not. We discuss the various properties of the orthogonal polynomials in
these problems and show that they are similar to those of the Bender-Dunne
polynomials in the one-body case. 
As a special case, we also discuss a self-dual system \cite{kuw1} 
in which case we can
obtain several more QES levels analytically.
Further, in this case we obtain a novel relation between the weight 
functions of the system.
Finally, in Sec. IV we summarize our conclusions and point out some of the
open problems. 

\section{Calogero-Marchioro type problem in $D$-dimensions}

The N-body Hamiltonian corresponding to the Calogero- Marchioro Model
in D-dimension for arbitrary potential $V$ can be written as \cite{kr}

\begin{equation}
H=-\frac{ \hbar^2}{2m}\sum_{i=1}^{N}{\bf {\nabla}}_i^2+g\sum_{i<j}^N \frac{
1}{r_{ij}^2}+ G\sum_{i<j,\ i,j\neq k}\frac{
{\bf {r}}_{ki}^2\cdot{\bf {r}}_{kj}^2}{r^2_{ki}r^2_{kj}} +V \left ( \sum_{i<j}
^Nr_{ij}^2\right )
\label{hcm}
\end{equation}
where ${\bf r}_{ij}= ({\bf r}_i-{\bf r}_j)$ and ${\bf r}_i$ is the 
D-dimensional position vector
for the i-th particle.

On substituting the ansatz
\begin{equation}
\psi = \left ( \prod_{i<j}{\bf {r}}_{ij}^2\right )^{\Lambda _D/2}\phi(\rho)
\label{psi}
\end{equation}
in the Schr$\ddot{o}$dinger equation ($H\psi = \epsilon \psi$) 
corresponding to the above system 
( with $\hbar = m =1$), it is easily shown that the equation satisfied 
by $\phi(\rho)$ is \cite{kr}
\begin{equation}
\phi^{\prime\prime}(\rho) + \frac{ 2\Gamma_D+1}{\rho}
\phi^\prime(\rho)-2V(\rho)\phi(\rho) +E \phi(\rho)=0 \, .
\label{phi}
\end{equation}
Here $E = 2\epsilon$ while 
\begin{eqnarray}
&& \rho^2 = \frac{ 1}{N}\sum_{i<j}^N r_{ij}^2 \nonumber \\  
&& \Gamma_D = \frac{1}{2}\left [D(N-1)-2 +\Lambda_D N(N-1) \right ]
\nonumber \\ 
&& \Lambda_D = \sqrt {G}= \frac{ 1}{2} \left [ \sqrt{(D-2)^2+4g
}-(D-2) \right ] \, .
\label{3}
\end{eqnarray}
We now consider two different forms of $V(\rho)$ and obtain QES solutions
in each case.

\subsection{Sextic Potential and QES problem}

Let us first consider $V(\rho)$ as given by
\begin{equation}
V(\rho) = \frac{1}{2} \left [B\rho^2 + C \rho^4 +H\rho^6+ \frac{ F}{\rho^2}
\right ] \, .
\label{v}
\end{equation}
In the special case when $C = 0 = H$, an infinite class of exact solutions
including the bosonic ground state have already been obtained \cite{kr,kha}.
 
On substituting 
\begin{equation}
\phi(\rho) = \rho^a {\exp {(-\alpha\rho^2-\beta\rho^4)}} \eta (\rho)
\label{pal}
\end{equation}
in Eq. (\ref{phi}) we obtain
\begin{eqnarray}
\eta^{\prime\prime} (\rho)+ \left ( \frac{ 2a+2 \Gamma _D
+1}{\rho}-[4\alpha\rho +8\beta\rho^3] \right ) 
\eta^{\prime} (\rho) +\hspace{1.4in} \nonumber \\  
  \left ( E-4 \alpha (a+ \Gamma _D +1)   
+ \rho^2 \left [ 4 \alpha ^2 -B -8\beta(a+ \Gamma _D +2) \right ] 
\right ) \eta(\rho) =0
\label{q}
\end{eqnarray}
where we have chosen,
$ \alpha = \frac{ C}{4\sqrt{H}};\ \ \ \beta = \frac{\sqrt{H}}{4} $ and $F=
a^2 +2a \Gamma _D$ \, . Thus $a$ is zero or nonzero depending on if $F$ is
zero or nonzero.
Finally, on substituting
\begin{equation}
\eta (\rho) = \sum_n \frac{ P_n (E) \rho^{2n}}{4^n n!(n+a+ \Gamma_D)!}
\label{pp1}
\end{equation}
in Eq. (\ref{q}) we obtain the recursion relation satisfied by
$P_n (E)$ as
\begin{eqnarray}
P_n (E)+ \left [ E- 4 \alpha \left ( 2n -1 +a+ \Gamma _D \right ) \right ]
P_{n-1} (E) 
+4(n-1)(n-1+a+ \Gamma _D)
\nonumber \\   \left [ 4 \alpha ^2 -B -8 \beta \left ( 2n +a+
\Gamma _D -2 \right ) \right ]P_{n-2} (E) =0 
\label{rec}
\end{eqnarray}
with initial conditions $P_{-1} =0 $ and $P_0=1$.
Using the well known theorem
\cite{chi,fgr},
`` the {\it necessary and sufficient} condition for a family of polynomials 
$\{P_n\}$ (with degree $P_n= n $) to form an orthogonal polynomial system is 
that $\{P_n\}$ satisfy a three-term recursion relation of the form
\begin{equation}
P_n(E) = \left ( A_n E +B_n \right ) P_{n-1}(E) + C_n P_{n-2}(E) \, , \ \ \ \  
n\ge 1
\label{r2}
\end{equation}
where the coefficients $A_n , B_n $ and $C_n$ are independent of $E$, 
$A_n \neq 0 , C_1 =0 , C_n \neq 0 $ for $n\ge 1$" , it then follows that
$\{P_n(E)\}$ for this problem forms an orthogonal set of polynomials with 
respect to some weight function, $\omega(E)$.

Let us write $B$ in the form 
$B= 4 \alpha ^2 - 8\beta (2J+a+ \Gamma
_D)$ where $J$ is any arbitrary number. We shall see that when J is a positive
integer, then this represents a QES system. In terms of $J$,
Eq. (\ref{rec}) then can be rewritten as 
\begin{eqnarray}
P_n (E)+ \left [ E- 4 \alpha \left ( 2n -1 +a+ \Gamma _D \right ) \right ]
P_{n-1} (E) \hspace{1in}\nonumber \\
-64\beta(n-1)(n-1+a+ \Gamma _D)  \left ( n-J-1 
 \right ) P_{n-2} (E) =0 
\label{rec1}
\end{eqnarray}
 from where it is clear that so long as  $J$ is positive integer, this 
recursion relation will reduce to
a two term recursion relation.
Thus, when $J$ is a positive integer, we have a QES system. These recursion
relations generate a set of orthogonal 
polynomials of which the first few are
\begin{eqnarray}
P_1 &=& -E + 4 \alpha (a+ \Gamma _D+1) \nonumber \\  
P_2 &=& E^2 - E \left [ 8 \alpha (a+ \Gamma _D+2) \right ] + 16 \alpha 
^2(a+\Gamma_D+1) (a+\Gamma_D+3) \nonumber \\  &&\hspace{2in} 
-64\beta (a+\Gamma_D+1) (J-1) \nonumber \\  
\label{pp}
\end{eqnarray}
It is easily seen that when $J$ is a positive integer, exact energy 
eigenvalues for the  
first $J$ levels are known. Further, when $J$ is a positive integer, these
polynomials exhibit 
the factorization property as given by  
\begin{equation}
P_{n+J}(E) = P_J(E) Q_n(E)
\label{fac}
\end{equation}
where the polynomial set $Q_n(E)$ correspond to the non-exact spectrum 
for this problem with $Q_0 (E) =1$. 
 For example, for $J=1$, $P_{n+1}$ will be factorized into $P_1$ and $Q_n$
and the corresponding QES energy level (which in this case is the ground 
state) is obtained by putting
$P_1 =0$ i. e. 
\begin{equation}
E_1= 4 \alpha (a+\Gamma_D+1) 
\label{e1}
\end{equation}

 Similarly for $J=2$, $P_{n+2}$ will be factorized into $P_2$ and $Q_n$ and
the corresponding
energy levels are,
\begin{eqnarray}
E_1 &=& 4 \alpha (a+\Gamma_D+2)  - 4 \sqrt{ \alpha ^2+4\beta (a+\Gamma_D+1) } 
\nonumber \\ 
E_2 &=& 4 \alpha (a+\Gamma_D+2)  + 4 \sqrt{ \alpha ^2+4\beta (a+\Gamma_D+1) }  
 \label{e2}
\end{eqnarray}

The quotient polynomials $Q_n (E)$ also form an orthogonal set as they 
satisfy the
recursion relation
\begin{eqnarray}
Q_n (E) + \left [ E- 4 \alpha \left \{ 2(n+J) -1 +a+ \Gamma _D \right \} \right ]
Q_{n-1} (E) \hspace{.8in} \nonumber \\  
-64\beta(n+J-1)(n+J-1+a+ \Gamma _D)  \left ( n-1 
 \right ) Q_{n-2} (E) =0 
\label{rec2}
\end{eqnarray}
 
with $Q_0 (E) =1$ and $Q_{-1} (E) =0$.

The square norm of both $P_n$ and $Q_n$ polynomials can be calculated from the
recursion relations in Eqs (\ref{rec1}) and (\ref{rec2}) respectively.
We obtain
\begin{eqnarray}
\gamma_n^{P} &=& 64\beta\prod_{k=1}^n k \left (k+a+\Gamma_D \right ) (J-n)
\nonumber \\   
\gamma_n^{Q} &=& 64\beta\prod_{k=1}^n (k+J) \left (k+J+a+\Gamma_D \right )  
\label{norm}
\end{eqnarray}

Note that a la the Bender-Dunne case \cite{bd} the norm for the $P_n$ 
polynomials vanish for
$n\geq J$ while it is positive definite for $n < J$. This is an alternative 
characterization  of the QES 
system. On the other hand, the  norm for the $Q_n $ polynomials is always 
non-vanishing and  
positive definite.

One can also calculate the weight functions $\omega_k$ for the set 
${P_n (E)}$ by using the
relations
\begin{equation}
\sum_{k=1}^J P_n (E_k)\omega_k = \delta_{n0} \, , \ n = 0,1,...,J-1 \, .
\label{wf}
\end{equation}
For example, for $J$ =2 we obtain
\begin{eqnarray}
\omega_1 &=& \frac{1}{2} + \frac{ \alpha}{2\sqrt{\alpha ^2+4\beta
(a+\Gamma_D+1) }} \nonumber \\  
\omega_2 &=& \frac{1}{2} - \frac{ \alpha}{2\sqrt{\alpha ^2+4\beta
(a+\Gamma_D+1) }}
\label{wt}
\end{eqnarray}

Notice that to the  leading order, the weight functions are 
inversely proportional to the particle number $N$ for a fixed $D$, 
 while they are inversely
proportional to the square root of the space dimension $D$, when the particle 
number $N$ remains unchanged. It may also be noted that both the weight 
functions $\omega_1$ and $\omega_2$ are positive definite. Actually  
one can prove on very general grounds that for any $J$, the weight
functions in this case will always be positive. As has been shown by 
Finkel et al. \cite{fgr}, if the three-term recursion relation is of the
form
\begin{equation}
P_{k+1} = (E-b_k)P_k -a_k P_{k-1} \, , \ k \ge 0
\label{pos}
\end{equation}
with $a_0 =0$ and $a_{n+1}$ =0, then (i) the weight functions are all 
positive (ii) the norm of the polynomials ($\gamma_n^p$) is positive
for $n <J$, 
provided if $b_k$ is real for $0 \le k \le n$ and  $a_k > 0$ for 
$1 \le k \le n$. Using Eq. (\ref{rec1})  
it is then easy to see that hence for any $J$, the weight
functions and $\gamma_{n}^p$ (for $n <J$)  will be positive. 

One can also calculate the moments 
$\mu_n =\int dE E^n\omega(E)$ of the weight functions. It is easily shown
that both the odd and the even moments are non-zero. Further, to the leading order
\begin{equation}
\mu_n = \left [ 4 \alpha (a+\Gamma_D+1) \right ]^n+ \cdots \, .
\label{gt} 
\end{equation}
 Thus to the leading order, the growth rate is proportional to
$N^{2n}$  for fixed $D$ and proportional to $D^n$ for fixed $N$.

Finally, let us discuss the consequences of the anti-isospectral 
transformation (also termed as duality transformation) \cite{kuw1} in the
context of our QES problem. It is easily seen that because of the duality
transformation ($x \longrightarrow ix$), the QES levels of the potential 
in Eq. (\ref{v}) are related to
that of a similar potential (say ${\hat V}$) where $C$ (and hence $\alpha$) 
has been replaced by
$-C$. In particular, if $J$ levels of the potential in Eq. (\ref{v}) are QES levels
with energy eigenvalues and eigenfunctions $E_k$ and $\psi_k$ respectively
($k = 1,2,...,J$), then the corresponding QES eigenvalues and eigenfunctions 
of ${\hat V}$
are given by
\begin{equation}
{\hat E}_k = - E_{J+1-k} \, , \ \ {\hat \psi}_k (x) = \psi_{J+1-k} (ix)
\label{d}
\end{equation}
For example, for $J$ =2, the QES energy eigenvalues of ${\hat V}$ are 
same as those given by Eq. (\ref{e2}) with the precise relationship
being ${\hat E}_{1,2} = -E_{2,1}$. The other properties of the dual potential
are also similarly related and can be easily worked out. 

\subsection{Self-dual Potential}

Let us now discuss the special case of $C =0$ in Eq. (\ref{v}) 
in which case we have a 
self-dual QES potential which has remarkably simple properties.
For example, the recursion relation (\ref{rec1}) now takes the simpler form
(note that now $\alpha = C/{4{\sqrt H}} = 0$)
\begin{equation}
P_n (E)+ E
P_{n-1} (E) 
-64\beta(n-1)(n-1+a+ \Gamma _D)  \left ( n-J-1 
 \right ) P_{n-2} (E) =0 
\label{rec3}
\end{equation}
and similarly the set {$Q_n$} satisfy Eq. ({\ref{rec2}) with $\alpha = 0$. 
As a result, both $P_n (E)$ and $Q_n (E)$ are eigenfunctions of parity. 
First few polynomials generated by this recursion relation are given by
$(s\equiv 1+a+\Gamma_D) $
\begin{eqnarray}
P_1 &=& -E \nonumber \\  
P_2 &=& E^2 -64\beta s(J-1) \nonumber \\ 
P_3 &=& -E^3+E \left [ 64\beta s(J-1)+128\beta(s+1)(J-2) \right ] \nonumber
\\ 
P_4 &=& E^4 -E^2 \left [ 64\beta s(J-1)+128\beta(s+1)(J-2) + 192\beta
(s+2)(J-3) \right ] \nonumber \\  
     && +(64\beta)^2 3s(s+2)(J-1)(J-3)
\label{spl}
\end{eqnarray}

From the duality relation (\ref{d}) it follows that for this self-dual
system, all the energy eigenvalues must be symmetrically distributed 
around $E =0$.
Further,
in this case QES eigen values can be analytically calculated very simply for 
$J = 1,2,...,5$.
For example, the eigen values for $J=3$ and $J=4$ are given as
\begin{eqnarray}
J=3\ \  : && \nonumber \\ 
&& E_1 = -8 \sqrt{2\beta(2s+1)} \nonumber \\ 
&& E_2 =0 \nonumber \\  
&& E_3 = 8\sqrt{2\beta(2s+1)}
\label{eng}
\end{eqnarray}
and
\begin{eqnarray}
J=4\ \  : && \nonumber \\ 
&& E_1 = - \sqrt{320\beta(s+1)+64\beta\sqrt{16s(s+2)+25}} \nonumber \\ 
&& E_2 = - \sqrt{320\beta(s+1)-64\beta\sqrt{16s(s+2)+25}} \nonumber \\ 
&& E_3 =  \sqrt{320\beta(s+1)-64\beta\sqrt{16s(s+2)+25}} \nonumber \\ 
&& E_4 = -\sqrt{320\beta(s+1)+64\beta\sqrt{16s(s+2)+25}}  
\label{eng1}
\end{eqnarray}
 
The square norms of the polynomial sets {$P_n$} and {$Q_n$}s are again given 
by Eq.
(\ref{norm}). Since the polynomials have definite parity the odd moments
vanish. The even moments of the weight functions in the leading order
are given by
\begin{equation}
\mu_{2n} \sim \left [ (1+a+\Gamma_D) \right ]^n+\cdots
\end{equation}
This means $\mu_{2n} $ is proportional to $N^{2n}$ for fixed $D$ and 
proportional to  $D^{n}$ for fixed $N$. It is worth noting that in the
limit $C =0$ (and hence $\alpha =0$), the leading term in the expansion of 
$\mu_n$ in Eq. (\ref{gt}) vanishes, and that is why the leading behavior
of
$\mu_n $ as function of $(1+a+\Gamma)$  when $C$ ( and hence $\alpha $) is 
non-zero is same as that of $\mu_{2n} $ when $ C $ ( and hence $ \alpha )=0$.  

The weight functions in this self dual QES model are related through a nice
relation 
\begin{equation}
\omega_k = \omega_{J+1-k}
\label{www}
\end{equation}
This can be proved explicitly using the relation in Eq. (\ref{wf}) as follows

We consider $ n$ to be odd in Eq. (\ref{wf}) , i.e.
\begin{equation} 
 \sum_k P_{2m+1}(E_k) \omega_k = 0\ \ \  \mbox{for}\ \  k= 1,2 ...,J
\label{01}
\end{equation}
Now since the $P_n(E)$ are of definite parity of $ E$, we can write
\begin{equation} 
P_{2m+1}(E_k) = \sum_{n=0}^m a_n E_k^{2n+1}\ \ ,\ \ m=0,1,2\cdots
\label{02}
\end{equation}
when $a_n$ are ($E$- independent) constant and $a_m
=-1$.

Putting (\ref{02})  in (\ref{01}) we obtain
\begin{equation}
\sum_{k=1}^J\sum_{n=0}^m a_n E_k^{2n+1} \omega _k =0 
\label{03}
\end{equation}
This implies that 
\begin{equation}
\sum_{k=1}^J E_k^{2m+1}\omega _k =0 
\label{04}
\end{equation}  
Now in Eq. (\ref{04}) we  use the duality relation  between the eigen values
of the self-dual system, i.e.
$ E_k = -E_{J+1-k}$ to obtain
\begin{equation}  
 \sum_{k=1}^J E_{J+1-k}^{2m+1} \omega _k =0
\label{05}
\end{equation}  
Next we make a change of variable $ k=J+1-j$ in Eq. (\ref{05}) to obtain
\begin{equation}  
\sum_{j=1}^J E_j^{2m+1} \omega _{J+1-j} =0
\label{06}
\end{equation}  
Subtracting Eq. (\ref{06}) from  Eq. (\ref{04})
\begin{equation}  
\sum_{k=1}^J E_k^{2m+1}[\omega _k-\omega _{J+1-k}] =0 
\label{07}
\end{equation}  
This is true for arbitrary $E_k$ and hence the result, $ \omega_k =
\omega_{J+1-k} $.

We now explicitly  calculate the weight functions for the case of $J=3$ and
$J=4$. We find,

\begin{eqnarray}
J=3 : \ \ &&\nonumber \\  
&&\omega_1 =\omega_3= \frac{ s}{2(2s+1)} \nonumber \\ 
&& \omega_2 = \frac{s+1}{2s+2}  
\label{wei3}
\end{eqnarray}
and for
\begin{eqnarray}
J=4 :\ \ && \nonumber \\ 
 &&\omega _1 =\omega_4=\frac{1}{4} \left [ 1- \frac{ 2s+5}{\sqrt{16s(s+2)+25}}
\right ] \nonumber \\  
 &&\omega _2 =\omega_3=\frac{1}{4} \left [ 1+ \frac{ 2s+5}{\sqrt{16s(s+2)+25}}
\right ]   
\label{wei4}
\end{eqnarray}
Not surprisingly, the weight function satisfy relation (\ref{www}) and
$\sum_{k=1}^J \omega_k =1$. 
\subsection{An unusual QES Problem}

Let us now discuss an unusual QES problem. To that end consider the following
potential
\begin{equation}
V(\rho) = {1\over 2} [B^2 \rho^2 - {C\over \rho} +{F\over \rho^2}]
\label{nqes}
\end{equation}
It is worth noting that for $C=0$ as well as for $B =0$, a class
of energy levels can be analytically obtained \cite{kr,kha}. For the special
case of $N=2$ i.e. 2-anyons in a magnetic field and experiencing Coulomb potential 
. This problem has been discussed by Myrheim et al. \cite{new}. We want 
to show that when both
of them are non-zero, it is an example of an unusual QES system in the sense
that for one parameter family of potentials (when there is a specific 
relation between $B, C$ and $F$), one always obtains one QES eigenvalue. 

On substituting
\begin{equation}
\phi(\rho) = \rho^a {\exp {(-B\rho^2/2)}} \eta (\rho)
\label{pal1}
\end{equation}
in Eq. (\ref{phi}) we obtain
\begin{eqnarray}
\eta^{\prime\prime} (\rho)+ \left [ \frac{ 2a+2 \Gamma _D
+1}{\rho}-(2B\rho) \right ] 
\eta^{\prime} (\rho) +\hspace{1.7in} \nonumber \\  
  \left [ E-2 (a+ \Gamma _D +1)B   
+ {C\over\rho} 
\right ] \eta(\rho) =0
\label{q1}
\end{eqnarray}
where we have chosen,
$F = a^2 +2a \Gamma _D$ \, .
Next we substitute 
\begin{equation}
\eta (\rho) = \sum_n { P_n (E) \rho^{n}}
\label{pp2}
\end{equation}
in Eq. (\ref{q1}) and obtain the recursion relation satisfied by
$P_n (E)$ as
\begin{eqnarray}
n(2a+2\Gamma_D +n)P_n (E)+ C P_{n-1} (E) + 
[ E- 2 (n -1 +a+ \Gamma _D) ]
P_{n-2} (E) = 0 
\label{rec4}
\end{eqnarray}
with initial conditions $P_{-1}=0$ and $P_0=1$.

From the recursion relation it follows that the QES levels are obtained in
case the coefficient of $P_{n-2} (E)$ vanishes, i.e. the energy of the QES
levels is given by $E = 2(a + \Gamma_D +n +1)$ where $n = 1,2,...$. 
Further, in that case B, C and F are not independent. 
For example, for $n =1$, the relation is $C^2 = 2 (2a +2\Gamma_D +1)B$ while
for $n=2$ the relation is $C^2 = 4(4a +4\Gamma_D +3)B$.
The QES level corresponds to an excited (ground) state depending on if 
$C > (<) 0$. On comparing Eq. (\ref{rec4}) with Eq. (\ref{r2}), we
conclude that the polynomials in this case do not correspond to 
orthogonal polynomials. 

\section{ Other N-body problems}
In this section we briefly discuss two other many body problems
and we will see that in both of these cases we have two QES problems which are
very similar to those discussed in the last section.

\subsection{ Novel Correlations for N- particles}
Recently, a class of exact solutions including the bosonic ground state has 
been 
obtained for a many body Hamiltonian in two dimensions such that the 
wave-functions have a 
novel correlation (of the form $X_{ij} = x_i y_j - x_j y_i$) built into 
them \cite{mbs}. 
Here we would like to show that 
QES states with novel correlations can also be obtained in this case. To that
purpose we start from the Hamiltonian 
\begin{eqnarray}
H &=& - \frac{ \hbar^2}{2m}\sum_i\nabla_i^2 + \frac{ \hbar^2}{2m} g_1\sum_{i,j}
\frac{ {\bf {r}}_{ij}^2}{X_{ij}^2} + \frac{\hbar^2}{2m} g_2\sum \frac{ {\bf {r}}
_j\cdot{\bf {r}}_k}{X_{ij}\cdot X_{ik}} \nonumber \\  
&& + \frac{ \hbar^2}{2m} \left [ 2 \left \{ B\sum_i r_i^2 + C(\sum_i r_i^2)^2
+H(\sum_ir_i^2)^3 + \frac{ F}{\sum_i r_i^2} \right \} \right ]
\label{h1}
\end{eqnarray}
On substituting  the ansatz 
\begin{equation}
\psi = \prod_{i<j}^N|X_{ij}|^g\phi({\rho})
 \end{equation}
in the Schr$\ddot{o}$dinger equation $H\psi=\epsilon \psi, \ (m=\hbar=1)$
one can show that $\phi$ satisfies the equation 
\begin{equation}
\phi^{\prime \prime }+\frac{ 2\Delta +1}{\rho}\phi^\prime 
+\left [E- B\rho^2 -C\rho^4-H\rho^6- \frac{ F}{\rho^2}\right ]\phi =0
 \label{nc}
\end{equation}
Where
\begin{eqnarray}
2\Delta +1 &=& 2N-1+2gN(N-1) \nonumber \\ 
\rho^2 =\sum_ir_i^2; && E = 2\epsilon ; \ \  g_1=g(g-1) \, , g_2 = g^2 \, 
\label{nc1}
\end{eqnarray}
On substituting
\begin{equation}
\phi(\rho) = \rho^a {\exp {(-\alpha\rho^2-\beta\rho^4)}} \eta (\rho)
\label{pal2}
\end{equation}
in Eq. (\ref{nc}) we obtain
\begin{eqnarray}
\eta^{\prime\prime} (\rho)+ \left [ \frac{ 2a+2 \Delta
+1}{\rho}-(4\alpha\rho +8\beta\rho^3) \right ] 
\eta^{\prime} (\rho) + \hspace{1.2in}\nonumber \\  
  \left [ E-4 \alpha (a+ \Delta +1)   
+ \rho^2 \left ( 4 \alpha ^2 -B -8\beta(a+ \Delta +2) \right ) 
\right ] \eta(\rho) =0
\label{q2}
\end{eqnarray}
where we have chosen,
$ \alpha = \frac{ C}{4\sqrt{H}};\ \ \ \beta = \frac{\sqrt{H}}{4} $ and $F=
a^2 +2a \Delta$ \, .
Now notice that this equation is similar to the corresponding 
equation in the last section and hence the rest of the analysis of that
section will
go through step by step in this case. Thus this is another example
of QES $N$-body problem in two dimensions where the polynomials 
satisfy most of the properties of the Bender-Dunne polynomials.
 Similarly, one can also consider
the admixture of the oscillator and Coulomb type potential and as in
the last section, again obtain similar QES solutions. 
In particular, in this case the polynomials do not form an orthogonal set.
\subsection{ N-body QES problem in 1-dimension}

Finally, we would like to briefly discuss the Calogero-Sutherland type 
$N$-body QES 
problems in one dimension \cite{ush1,mrt}. 

 The Hamiltonian corresponding to the N-body QES problem in 1-dimension is
given by 
\begin{eqnarray}
H= -\frac{ \hbar^2}{2m} \sum_i\nabla_i^2 +\sum_{i<j } \frac{
g}{(x_i-x_j)^2} + B\sum_{i} x_i^2 + C [\sum x_i^2]^2 \nonumber \\  
H [ \sum x_i^2]^3 + \frac{ F}{\sum x_i^2 }
\label{h2}
\end{eqnarray}

We substitute the standard ansatz for the wave function $\psi$,
\begin{equation}
\psi = Z^{\lambda+ \frac{1}{2}} \phi(\rho)
\end{equation}
where $ Z=\prod_{i<j} (x_i-x_j),\  \rho^2=\sum x_i^2,\  
\lambda= \frac{1}{2}\sqrt{1+4g}\ $
in $H\psi =\epsilon \psi \ (\hbar=m=1) $ and obtain the $\phi$-equation
\begin{equation}
\phi^{\prime \prime } + \frac{ 2\Delta+1}{\rho}\phi^\prime + \left [
E-B\rho^2-C\rho^4 -H\rho^6- \frac{F}{\rho^2} \right ]\phi =0
\end{equation}
where $E = 2\epsilon$ and  
\begin{eqnarray}
2\Delta+1 = N-1+N(N-1)\lambda \nonumber \\ 
\end{eqnarray}
This equation is very similar to the Eq. (\ref{nc}) and hence the rest of the
analysis goes through as in the last section, i.e. the 
polynomials of the QES problem indeed form an orthogonal set and 
share most of the properties of the Bender-Dunne polynomials.
Similarly, we can show that even in this case the other QES solution 
corresponding to the admixture of the oscillator and Coulomb potential
also exists, and in this case, the polynomials do not form an orthogonal
set. 

\section{Summary and Open Problems}

In this paper, we have obtained QES solutions of a number of $N$-body 
problems in one,two and higher dimensions. We have shown that with sextic
type potential, in all these cases one has a set of orthogonal polynomials
which satisfy almost all the properties as satisfied by the Bender-Dunne 
polynomials. However, the hidden algebra is not clear in these cases. In this
context, recall that for the case of one particle in one dimension 
experiencing the sextic
potential, the underlying symmetry algebra is $Sl(2)$. In particular, in that
case one  
can express the Hamiltonian in terms of the quadratic (and linear) generators
of $Sl(2)$. Further, it will be nice to discover some other QES many-body 
problems. Finally, it would be most interesting if one can find some 
application of these QES $N$-body problems. In this context it is
worth noting that one application has already been found by one of the present
author (AK) and Jatkar \cite{jk}. In particular, they have considered the 
model discussed in Sec. IIIB (with F =0) and have shown that the square of the 
ground state wave function of this model is related to the matrix model 
corresponding to branched polymers. It will really be interesting, if one
can find some other application, of the N-body QES problems in two and 
higher dimensions.


\newpage
\begin{references}
\bibitem{ush} A. Ushveridze, {\it Quasi-Exactly Solvable Models
in Quantum Mechanics}, Inst. of Physics Publishing, Bristol, (1994). 
\bibitem{ush1} A. G. Ushveridze, Mod. Phy. lett {\bf A6} (1991) 977.
\bibitem{mrt} A. Minzoni, M Rosenbaum and A. Turbiner, hep-th/9606092.
\bibitem{jat}D. P. Jatkar, C. Nagaraja Kumar and A. Khare, Phys. Lett.
{\bf A 142} (1989) 200.
\bibitem{bpm} A. Khare and B.P.Mandal, Physics/9709043 (To appear in
Phys. Lett. A, 1998)
\bibitem{bd} C. M. Bender and G. V. Dunne, J. Math. Phys. {\bf 37} (1996) 6.
\bibitem{bpm1} A. Khare and B. P. Mandal , quant-ph/9711001 ( To appear
in J. Math Phys. 1998)
\bibitem{cm} F. Calogero and C. Marchioro , J. Math Phys. {\bf 14} (1973), 182.
\bibitem{kr} A. Khare and K. Ray, Phys. Lett. {\bf A230} (1997) 139.
\bibitem{mbs} M.V.N. Murthy, R.K. Bhaduri and D. Sen, Phys. Rev. Lett.
{\bf 76} (1996) 4103; 
R. K. Bhaduri, A. Khare, J. Law, M. V. N. Murthy and D. Sen
, J. Phys. A : Math. Gen. {\bf 30} (1997) 2557.
\bibitem{cal} F. Calogero, J. Math. Phys. {\bf 10} (1969) 2191, 2197, 
ibid {\bf 12} (1971) 419.
\bibitem{kuw1} A. Krajewska, A. Ushveridze and Z. Walczak,
Mod. Phys. Lett. {\bf A 12} (1997) 1225.
\bibitem{kha} A. Khare, cond-mat/9712133.
\bibitem{chi} T.S. Chihara, {\it An Introduction to Orthogonal Polynomials }
, Gordon and Breach, New York, (1975); A. Erdelyi, W. Magnus, F. Oberhettinger
and F. G. Tricomi, {\it Higher Transcendental Functions}, Vol.II, McGraw-Hill
, New York, (1953).
\bibitem{fgr}  F. Finkel, A. Gonzalez-Lopez And M. A. Rodriguez,
J. Math. Phys. {\bf 37} (1996) 3954.
\bibitem{kuw} A. Krajewska, A. Ushveridze and Z. Walczak,
Mod. Phys. Lett. {\bf A 12} (1997) 1131.
\bibitem{new} J. Myrheim, E. Halvorsen and A. Vercin, Phys. Lett. {\bf B
278} (1992), 171.
\bibitem{jk} D. Jatkar and A. Khare, Int. J. Mod. Phys. {\bf A11} (1996) 1357.
\end{references}
\end{document}